# Iterative Time-Varying Filter Algorithm Based on Discrete Linear Chirp Transform

Osama A. S. Alkishriwo[1*], Ali A. Elghariani[2], Aydin Akan[3]

[1] alkishriewo@yahoo.com, [2] elghariani_99@yahoo.com, [3] akan@istanbul.edu.tr

[1, 2] Department of Electrical and Electronic Eng., College of Eng., University of Tripoli, Libya
[3] Department of Electrical and Electronic Engineering, Istanbul University, Turkey
*Corresponding author email: alkishriewo@yahoo.com

**ABSTRACT**

Denoising of broadband non--stationary signals is a challenging problem in communication systems. In this paper, we introduce a time-varying filter algorithm based on the discrete linear chirp transform (DLCT), which provides local signal decomposition in terms of linear chirps. The method relies on the ability of the DLCT for providing a sparse representation to a wide class of broadband signals. The performance of the proposed algorithm is compared with the discrete fractional Fourier transform (DFrFT) filtering algorithm. Simulation results show that the DLCT algorithm provides better performance than the DFrFT algorithm and consequently achieves high quality filtering.

## 1  Introduction

Non–stationary relates to the time--dependence of the statistics of a random process. As such, non-stationary signals display either time–varying mean, variance or evolving spectra, or a combination of some or all of these. It is thus why more sophisticated filtering approaches are needed for the representation and processing of non–stationary signals.

Noise is an unwanted component which acts as a source of error in the signal analysis and should be suppressed before data processing and interpretation. In many cases, therefore, noise mitigation is essential in order to extract useful information from the signal contaminated in noise.

The Fourier transform is one conventional signal denoising technique. However, due to the nonlinear and non–stationary properties of signals, this method is limited in the denoising capability to this kind of signals [1]. To overcome this shortcoming, several new approaches have been proposed in literature. In [2, 3] the least mean square adaptive algorithms are presented, but these algorithm are not able to track the rapidly varying non–stationary signals. Moreover, the wavelet transform (WT) based methods are widely used because of their abilities to remove Gaussian noise. However, the performance of the WT-based denoising methods depends on their selected mother wavelets [4].

To capture the variability with time of the non–stationary signal parameters, it is necessary to consider extensions of the Fourier–based representations capable of providing instantaneous–frequency information for multi–component signals. Although this can be achieved by considering polynomial–phase transforms [5], second-order polynomial transforms [6, 7] are preferable due to computational viability.

In [7] the discrete linear chirp transform (DLCT) is introduced to represent a signal as a combination of linear chirps. The DLCT is an extension of the discrete Fourier transform (DFT) and provides a parametric modelling of the instantaneous frequencies of the components. Rather than joint time–frequency, the DLCT is a joint chirp–rate frequency transformation. It can be implemented efficiently using the fast Fourier transform (FFT) algorithm [7].

In this paper, a time–varying filtering algorithm based on the DLCT is proposed. The algorithm relies on the ability of the DLCT to decompose a signal iteratively into its components locally. Each of these components is filtered separately and then synthesized with the other filtered components to estimate the desired signal. Since each segment of the signal has different components with different bandwidths, the filter has to be time–varying. The performance of the proposed algorithm is compared with the discrete fractional Fourier transform (DFrFT) based filtering method [8, 9]. Simulation results of the proposed method show better performance on the denoising in comparing with major denoising schemes based on the DFrFT.

The paper is organized as follows. Section 2 shows how to obtain the DLCT and presents some of its properties. In section 3, we introduce the DLCT filtering algorithm. Simulation results are given in section 4, where we compare the DFrFT with the DLCT. In particular, we consider which of these two transforms is more efficient in transforming a non–sparse signal into a sparse–signal in time or in frequency, the resolution of the transforms, and the computational time required. Then, we evaluate the capability of the DLCT in filtering nonstationary signals. Finally, conclusions are summarized in section 5.

## 2 The Discrete Linear Chirp Transform (DLCT)

For a discrete–time signal $x(n)$, $0 \leq n \leq N-1$, its discrete linear chirp transform (DLCT) and its inverse are given by [7]

$$x(k,m) = \sum_{n=0}^{N-1} x(n) \, exp\left(-j\frac{2\pi}{N}(c\,m\,n^2 + k\,n)\right) \tag{1}$$

$$x(n) = \sum_{m=-L/2}^{L/2-1} \sum_{k=0}^{N-1} \frac{X(k,m)}{LN} \, exp\left(j\frac{2\pi}{N}(c\,m\,n^2 + k\,n)\right) \tag{2}$$

where $c$ is the resolution of the transform, $L$ and $N$ are the number of samples in the chirp–rate, and in the frequency domain, respectively. The DLCT is a joint chirp–rate frequency transformation that generalizes the discrete Fourier transform (DFT): indeed

$$X(k,m) = \frac{1}{N} X(k) \odot DFT \left\{ exp\left(j\frac{2\pi}{N} c\, m\right) \right\} \qquad (3)$$

where $\odot$ is the circular convolution. Hence, If $m = 0$, then $X(k, 0)$ is the DFT of $x(n)$. Thus, the DLCT can be used to represent signals that are locally combinations of sinusoids, chirps or both.

## 3   The Proposed DLCT–Based Filtering Algorithm

For a signal $x(n)$ we can identify from its DLCT $X(k,m)$ the number of components $Q$, the chirp–rates $\beta_i = c\, m_i$, and frequency parameters $k_i$. The energy concentration is indicated by the peak values of $|X(k,m)|^2$ as a function of $k$ and $m$. Considering the region in the joint chirp–rate frequency plane where these peak values occur, we should find the values of the chirp–rates and frequencies that can be used to approximate the given signal locally as a sum of linear chirp components

$$x(n) = \sum_{i=1}^{Q} x_i(n) \qquad (4)$$

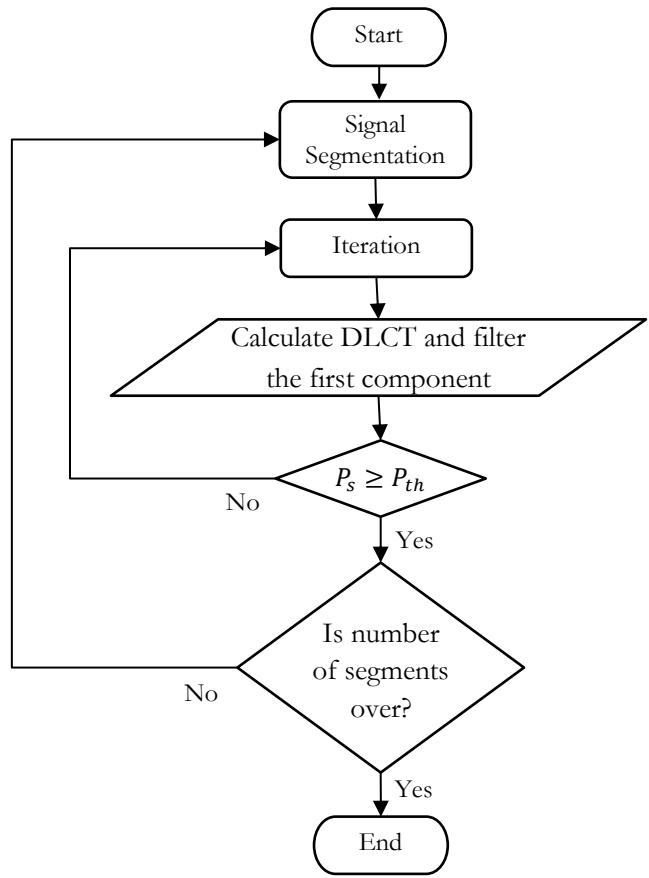

**Figure 1:** Filtering algorithm flowchart.

If we consider the case of a linear chirp contaminated in white Gaussian noise, filtering it directly without processing will permit for a large amount of noise to stay with the chirp since it has a broad bandwidth. However, if we implement the filtering process in the DLCT domain where the linear chirp has narrow bandwidth, then most of the noise will be removed.

The proposed DLCT filtering algorithm is performed over multiple iterations. Flowchart given in Figure 1 shows the step by step operation of the algorithm. The stopping criterion for the sub-iteration can be set manually by the expected number of components or adaptively based on a threshold $P_{th}$ of the remaining energy $P_s$ in the residual.

## 4  Simulation Results and Discussion

To evaluate the performance of the proposed algorithm, a simulation is performed to observe and compare the mean absolute error of the filtered signals with the DFrFT filtering algorithm. Both algorithms are applied to a synthetic signal as well as a real world signal. The synthetic signal is generated as follows,

$$x(n) = exp\left(j\frac{\pi}{256}(0.1\,n^2 + 10\,n)\right)$$

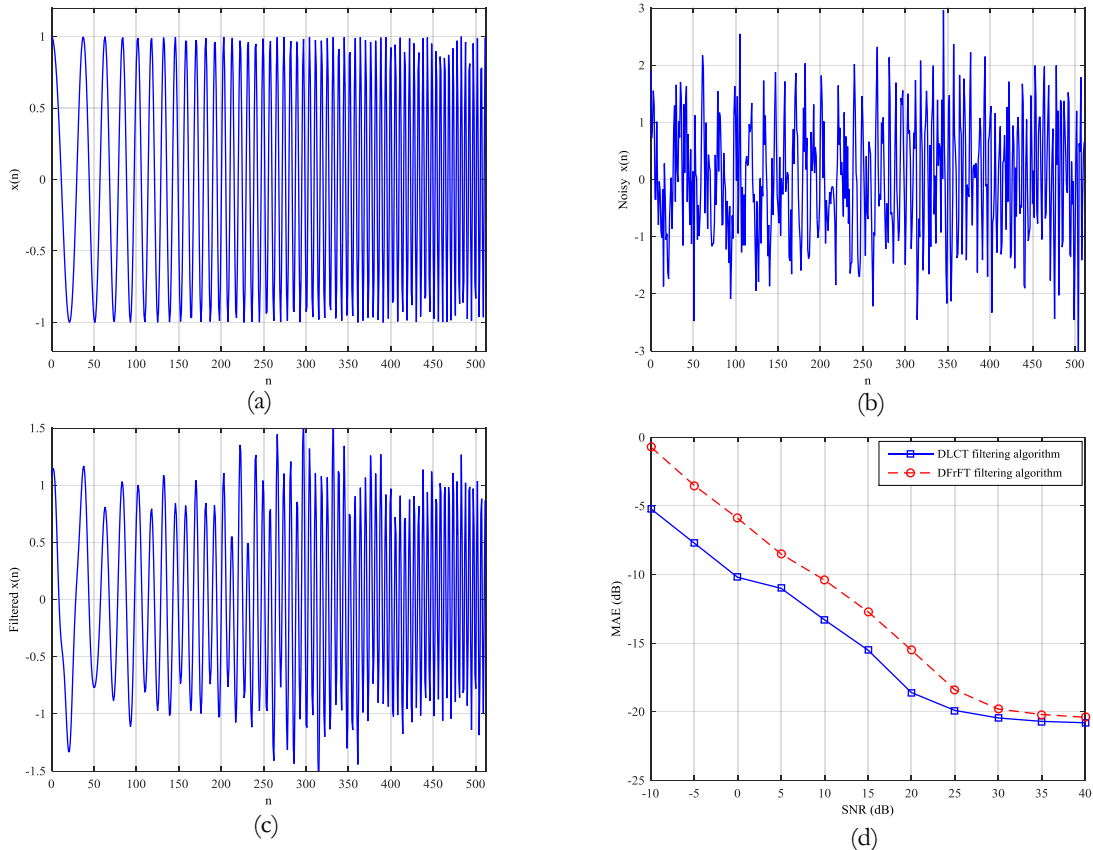

**Figure 2:** Linear chirp signal, (a) the chirp signal, (b) the noisy chirp signal with SNR= 0 dB, (c) the filtered chirp signal using DLCT filtering algorithm, and (d) mean absolute error for the two algorithms.

If the noiseless signal is denoted by $x(n)$ and $\hat{x}(n)$ is the filtered signal, then the mean absolute error can be calculated as follows,

$$MAE = \frac{1}{N} \sum_{n=0}^{N-1} |x(n) - \hat{x}(n)| \qquad (5)$$

Figures 2(a), (b), and (c) show the chirp signal, the noisy chirp signal, and the filtered chirp signal using the DLCT filtering algorithm, respectively. In Figure 2(d), we provide the mean absolute error (MAE) for the two algorithms against signal–to–noise ratio (SNR). It can be seen that the DLCT filtering algorithm outperforms the performance of the DFrFT filtering algorithm, especially at low SNR. For instance, at 0 dB SNR, DLCT has a 5 dB improvement, while at 25 dB SNR, it has only about a 1 dB improvement over DFrFT.

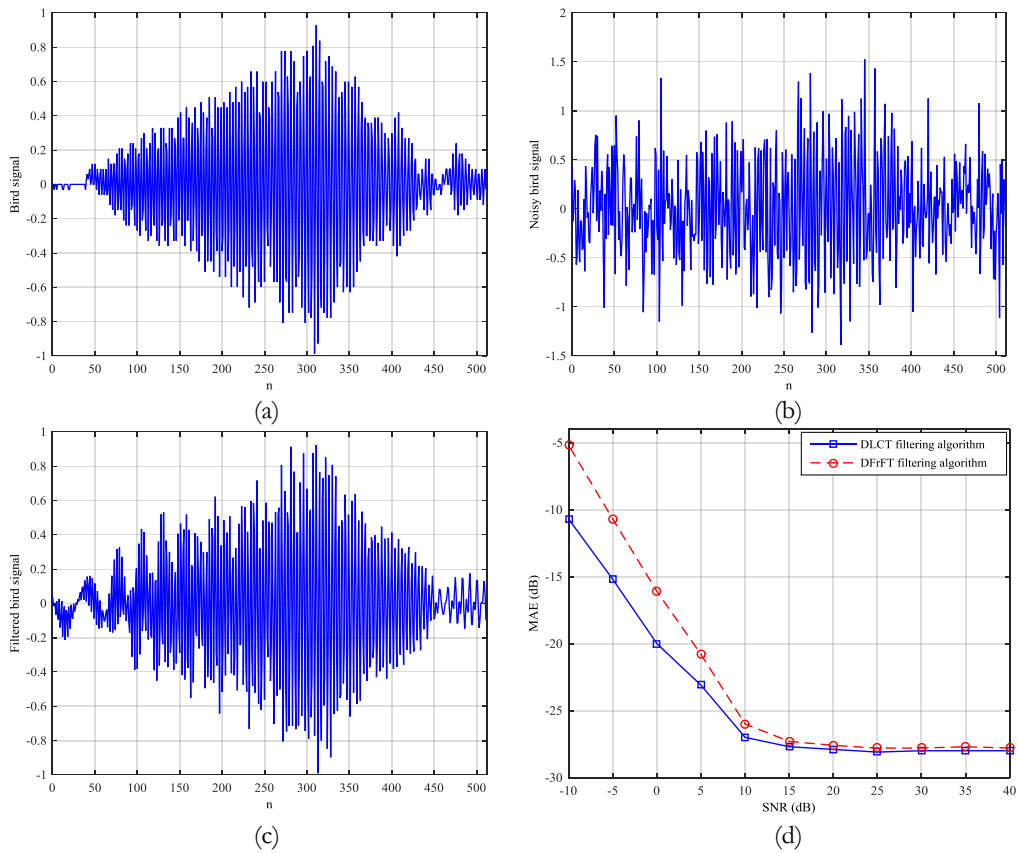

**Figure 3:** Real–world signal: (a) the bird chirping signal, (b) the noisy bird chirping signal with SNR= 0 dB, (c) the filtered bird chirping signal using DLCT filtering algorithm, and (d) the mean absolute error for the two algorithms.

To quantify the MAE improvement, a real–world signal (Bird chirping signal) with varying noise level is also simulated. The noiseless and noisy bird chirping signals are presented in Figures. 3(a) and (b). The denoised bird chirping signal based on the DLCT filtering algorithm is shown in Figure 3(c) at SNR=0 dB. Similar to the previous case, Figure 3(d) depicts the MAE for the two algorithms as a function of the input SNR, where input SNR is varied from −10 dB (severely poor SNR) to 40 dB (high SNR).

The DLCT filtering algorithm greatly enhances the MAE by simply employing few iterations in the simulation. This gain occurs because the DLCT algorithm gives sparser signals than the DFrFT algorithm.

## 5   Conclusion

In this paper, an iterative time–varying filtering algorithm based on the DLCT transform is proposed. The method exploits sparsity structure of signals to improve denoising performance of non-stationary signals. The performance of the proposed algorithm is analysed and compared with the performance of the (DFrFT) based method. Simulation results show that the DLCT filtering algorithm gives lower mean absolute error results than the DFrFT filtering algorithm in both synthetic and real–world signals. For instance, at SNR=0 dB, the proposed algorithm has a 5 dB improvement over the DFrFT filtering algorithm.

## References

[1]   M. Alfaouri and K. Daqrouq, "ECG Signal Denoising By Wavelet Transform Thresholding ," *American Journal of Applied Sciences*, vol. 5, no. 3, pp. 276-281, 2008.
[2]   V. Almenar and A. Albiol, "A new adaptive scheme for ECG enhancement," *Signal Processing*, Vol. 75, no. 3, pp. 253-263, Jun. 1999.
[3]   L. Durak and  S. Aldirmaz, "Adaptive fractional Fourier domain filtering,"  *Signal Processing*, vol. 90, no. 4, pp. 1188-1196, Apr. 2010.
[4]   J. P. Amezquita-Sanchez and H. Adeli, "A new MUSIC-empirical wavelet transform methodology for time-frequency analysis of noisy nonlinear and non-stationary signals," Digital *Signal Processing*, vol.45, pp. 55–68, Oct 2015.
[5]   S. Peleg and B. Friedlander, "The discrete polynomial-phase transform,"  *IEEE Transactions on Signal Processing*, vol. 43, no. 8, pp. 1901-1914, Aug. 1995.
[6]   C. Candan, M. Kutay, and H. Ozaktas, "The discrete fractional Fourier transform," *IEEE Trans. on Signal Processing*, vol. 48, no. 5, pp. 1329-1337, May 2000.
[7]   O. A. Alkishriwo and L. F. Chaparro, "A Discrete Linear Chirp Transform (DLCT) for Data Compression,"  *in Proc. of the IEEE International Conf. on Information Science, Signal Processing and their Applications*, Montreal, Canada, Jul. 2012, pp. 1283-1288.
 [8]   O. A. Alkishriwo, L. F. Chaparro, and A. Akan, "Signal separation in the Wigner distribution using fractional Fourier transform," *European Signal Processing Conf., EUSIPCO*,  Spain, Sep. 2011, pp. 1879-1883.
[9]    P. Kumar and S. Kansal, "Noise removal in speech signal using fractional Fourier transform, " *2017 International Conference on Information, Communication, Instrumentation and Control, ICICIC,* Indore, India,  Aug. 2017.